\documentclass[conference]{IEEEtran}
\usepackage{cite}

\ifCLASSINFOpdf
   \usepackage[pdftex]{graphicx}
   \DeclareGraphicsExtensions{.pdf}
\else
\fi
\begin{document}
%
\title{A Wideband, Four-Element, All-Digital Beamforming System for Dense Aperture Arrays in Radio Astronomy}





%
\author{\IEEEauthorblockN{Richard P. Armstrong\IEEEauthorrefmark{1}\IEEEauthorrefmark{3},
Kristian Zarb Adami\IEEEauthorrefmark{1}
and Mike E. Jones\IEEEauthorrefmark{1}
}
\IEEEauthorblockA{\IEEEauthorrefmark{1}University of Oxford Astrophysics,
Department of Physics,
Denys Wilkinson Building,
Keble Road,
Oxford, OX1 3RH}
\IEEEauthorrefmark{3}corresponding author: richard.armstrong@physics.ox.ac.uk
}

\maketitle

\begin{abstract}
Densely-packed, all-digital aperture arrays form a key area of technology development required for the Square Kilometre Array (SKA) radio telescope.  The design of real-time signal processing systems for digital aperture arrays is currently a central challenge in pathfinder projects worldwide. We describe an hierarchical, frequency-domain beamforming architecture for synthesising a sky beam from the wideband antenna feeds of digital aperture arrays.
\end{abstract}


%

\section{Introduction}

Densely-packed, all-digital aperture arrays form a key area of technology development required for the Square Kilometre Array (SKA) radio telescope. The design of real-time signal processing systems for aperture arrays is certainly one of the most challenging tasks in the design of next-generation instruments. In this paper, we describe the design and implementation of an hierarchical beamforming architecture for wideband radio telescopes.\\

This document is organised as follows. Chapter I is an introduction to the problem domain. Chapter II describes the theory and process of `beamforming' for aperture arrays used in radio astronomy. Chapter III describes the design and implementation of a frequency-domain beamformer. Chapter IV  describes calibration procedures for the array, and is followed by results and conclusions.\\

\subsection{Aperture Phased Arrays}
\begin{figure}
\begin{center}
\includegraphics[width=70mm]{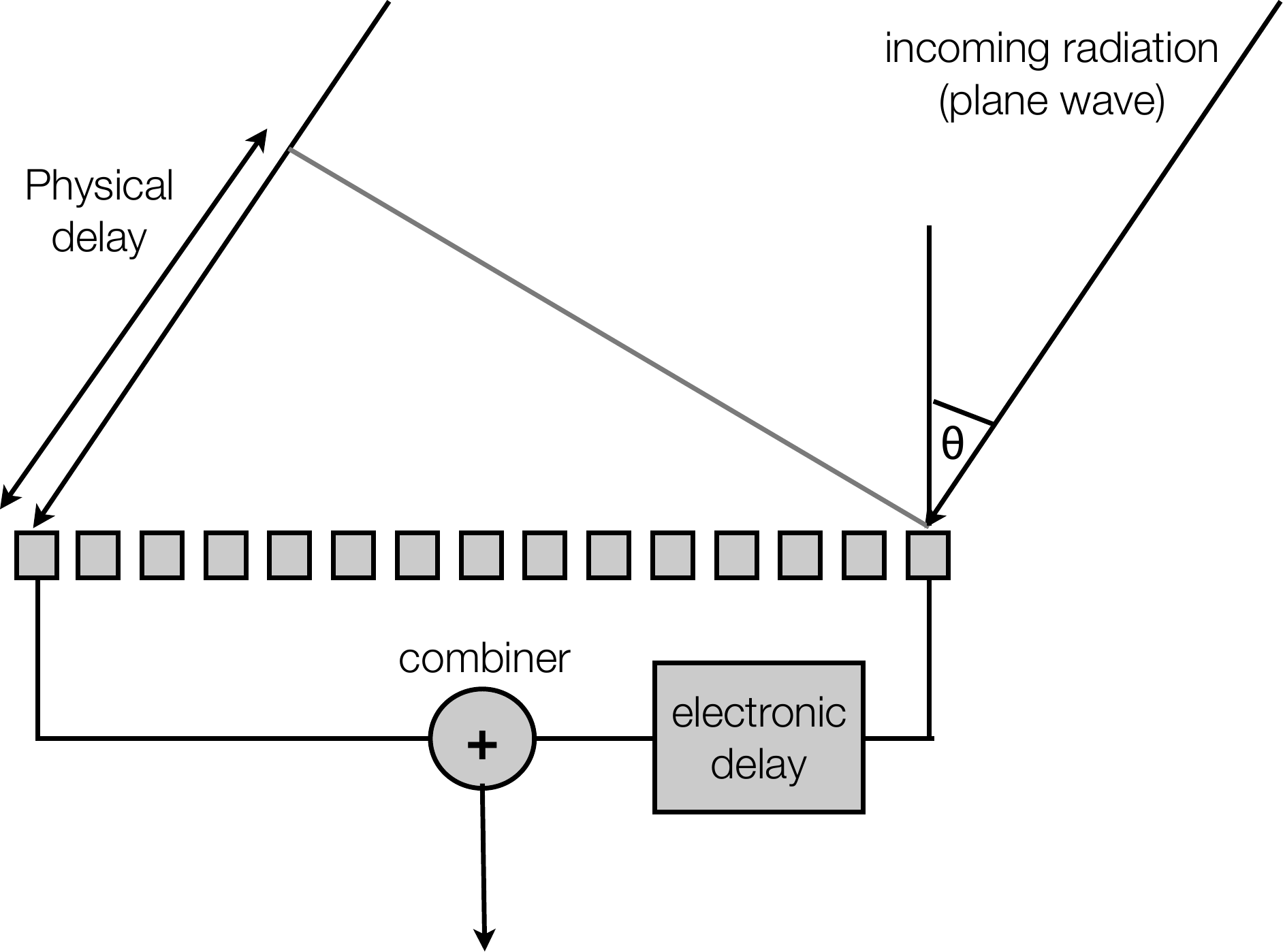}
\caption
	{An illustration of the electronic processing of signals received from an array of antennas for concentrating the signal from a certain direction. This is the basic principle of the phased aperture array: different delay configurations allow beams to be formed in different directions of \(\theta\)} 
\label{BeamformingIllustration}
\end{center}
\end{figure}

An aperture phased array is a collection of fixed, non-directional antennas whose outputs are combined to synthesise an effective directional antenna. This synthesised antenna is pointed electronically; significantly the physical component antennas do not ever move. The electronic combination of signals with beamforming techniques performs the same spatial filtering task as a parabolic dish would in a traditional radio telescope dish. Figure \ref{BeamformingIllustration} illustrates the basic principle of this operation.\\

The flexibility, survey speed, system-level autonomy and scalability provided by antenna arrays make them extremely attractive for next-generation radio instruments, for use, in our application, in radio astronomy, but also applicable to other wideband detection and communication instruments. Future radio instruments will certainly be dependent on the availability of high-performance signal processing systems such as these. Since aperture arrays are seen as a critical antenna technology to achieve the required sky survey speed in future radio telescopes (see, for example, \cite{2006SKA.memo.81} and \cite{2007skads.memo.FeA}), the development of a scalable signal processing system is of great importance.\\

\section{Beamforming}

\subsection{Digital, Frequency-Domain Beamforming}
The process of beam forming, shaping and summing of tile elements may be performed in either the digital or analogue domains. Digital beamformers are distinguished from analogue and partially analogue beamformers by signal digitisation directly after initial RF amplification, followed by processing entirely in the digital domain. The all-digital approach offers maximum flexibility in manipulation of incoming signals both in beamforming and in calibration of the wideband signals. However, there are many tradeoffs in this space, and different applications will dictate eventual implementation.\\

Beamforming techniques may be generally categorised as time-domain or frequency domain. For broadband applications, sub-banded frequency-domain techniques may be seen as superior due to their computational advantage if fine sample interpolation operation is implemented (see, for example, \cite{1027910}) and the ability to calibrate band-pass characteristics of the analogue system (see \cite{priv:2009}) by applying a phase shift to each sub-band in the transform domain. For these reasons, we choose to perform the beamfoming in Fourier space.\\



\subsection{Signal Processing and Dataflow Requirements for Wideband Phased Aperture Arrays}
In this section we describe the signal processing and dataflow requirements of the narrowband phase shift beamforming architecture. \\

The sampled broadband signal from each antenna is broken up into many narrow frequency bands. Each band is independently processed (delayed and combined with the signals from other antennas) with a narrowband phase-shift beamformer. The results of these narrowband beamformers are combined to form a broadband beam in the required spatial direction. This operation is repeated for each beam that is formed. Channelisation or `frequency binning' of large-bandwidth input signals for beamforming can be performed with various techniques.\\

\begin{figure}
\includegraphics[width=80mm]{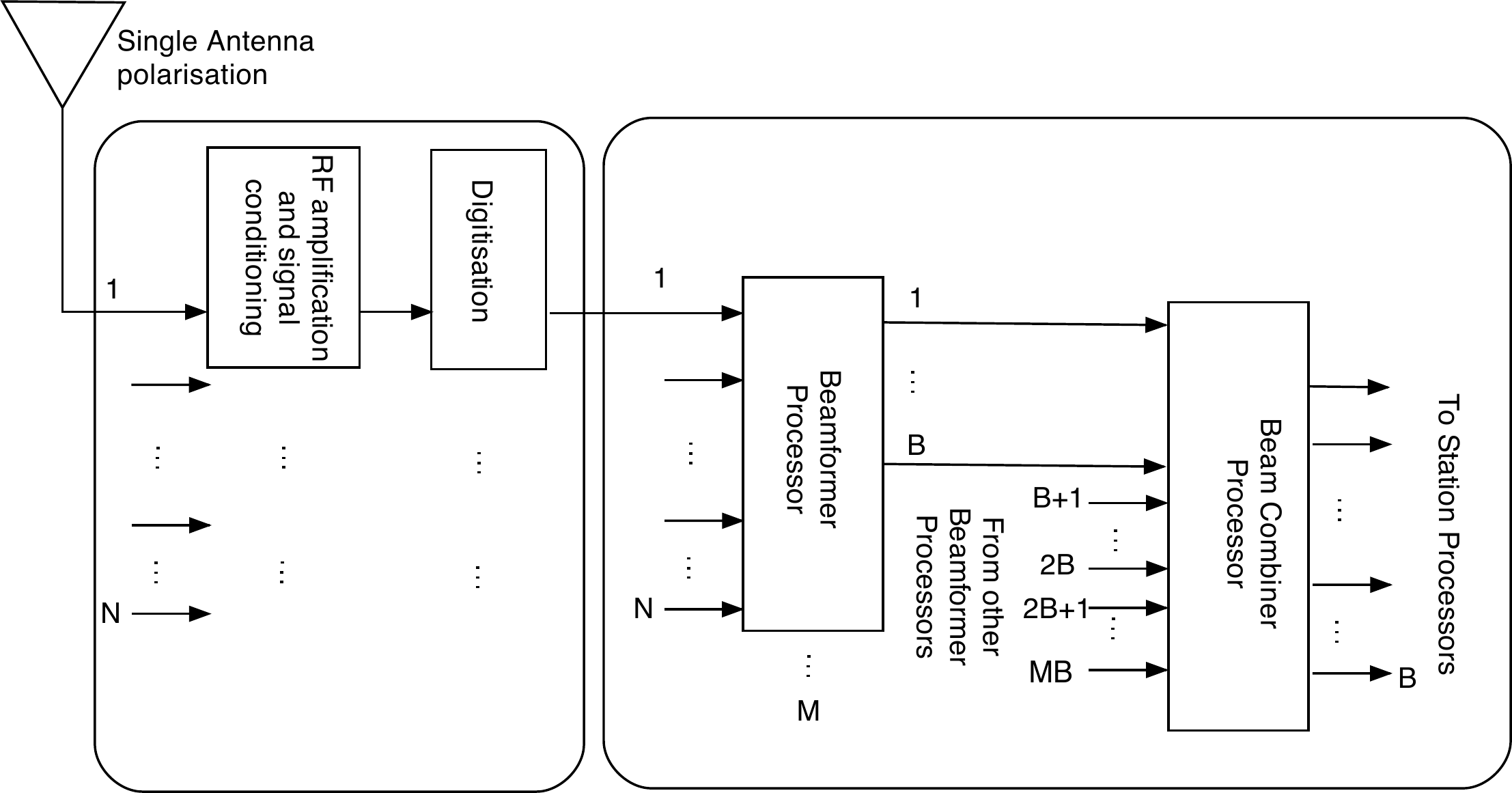}
\caption[System Signal Processing Flow for Digital Aperture Array Beamforming]
	{System Signal Processing Flow for Digital Aperture Array Beamforming at Tile Level} 
\label{SPFlowAA}
\end{figure}

Figure \ref{SPFlowAA} shows the signal processing flow for the phase-shift beamforming architecture. N single-polarisation antenna signals are digitised then combined with phase correction factors to form B beams in the beamformer processor. These beams are combined in a beam-combiner processor\footnote{The beam-combiner processor may be a further hierarchy of processors}. The resultant `station beams' (which are wideband spectra) are then sent to a central correlator for interferometric processing.\\

We now analyse the dataflow requirements: the analogue input signal ranges from 0.5GHz to 0.7GHz (a bandwidth of 0.2GHz), and is sampled with $q$ bits of precision, which results in a digital input bandwidth from each antenna polarisation of $q$ *0.4Gbps. This means that the first-stage beamformer processor will have a raw input data-rate of $Nq*0.4$ Gbps and an output data rate of $Bqs$*0.4 Gbps, where $s$ is the bitwidth scaling factor of the beamformer.\\

Note that there is a strong distinction between high-performance, real-time processing which must be performed at the input data-rate and more computationally complex co-efficient calculation which may be performed at a much reduced rate. In this paper, we concentrate on the former; this is an implementation of the high-performance processing system. We allow the steering and correction coefficients to be updated at a lower rate, dictated by the speed with which the array must be able to scan the sky.\\

\section{An Hierarchical, Frequency-Domain Beamforming Architecture}
In this section, we describe the design and implementation of an FPGA-based digital system for the frequency sub-band phase-shift beamforming architecture.\\

\begin{figure}
\begin{center}
\includegraphics[width=90mm]{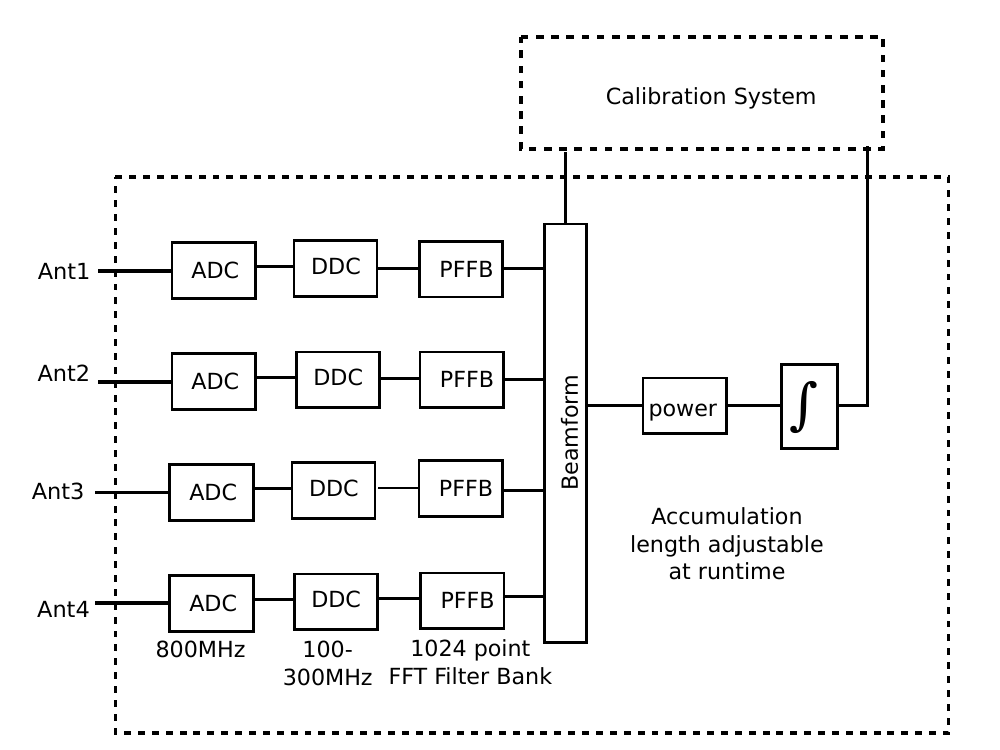}
\caption{An architectural drawing of the 4-element beamformer and calibration system} 
\label{2pad_arch}
\end{center}
\end{figure}

Figure \ref{2pad_arch} is an architectural schematic of the 4-element beamformer digital system. The Analogue to Digital Converter (ADC) samples the incoming antenna signal at 800MSa/s with 8 bits of precision. This 400MHz bandwidth signal is immediately Digitally Down-Converted (DDC) by mixing with an complex sinusoid at 3/4 of the ADC clock rate to yield a base-banded 500MHz to 700MHz input signal. Channelisation is the next step in the processing pipeline: an efficient technique is the Polyphase FFT\footnote{Fast Fourier Transform (FFT)} Filter Bank (PFFB) channeliser, which improves frequency isolation of the FFT operation (see for example \cite{2008PASP..120.1207P}) and has shown to be efficient for large numbers of equally spaced channels \cite{Pucker01}. The beamformer applies steering and correction coefficients to each signal stream and then performs beam summation. We then accumulate the power sum of the raw voltage beam signal for a run-time configurable length of time.\\

We have implemented this design with signal processing libraries and hardware designed by the Collaboration for Astronomy Signal Processing Research (CASPER). We are indebted to all collaborators of this group for the support they have provided with hardware and libraries.\\

The FPGA-based design uses embedded multipliers in a Xilinx Virtex II Pro device. Since these fixed hardware blocks are natively 18-bit, we preferentially use 18-bit multiplications and datapaths throughout the design.

\section{Calibration at Aperture Array Tile Level}

Calibration of individual elements of a dense aperture array at the tile level is fundamentally different from traditional radio telescope calibration for two reasons. Firstly, individual antennas `see' the entire observable sky; the dish-centric concept of Field of View (FoV) at the antenna level is not valid. Secondly, each individual element does not have the sensitivity to detect astronomical sources.\\

Thus we must approach the calibration in other ways. Perhaps the obvious first suggestion would be to fully constrain the analogue system. Certainly, this is done to the first degree. We refer the reader to \cite{priv:2009} for discussions about analogue signal chains, including S-parameter propagation, variation due to temperature, component anisotropy and radio-frequency interference (RFI) effects. That is, a typical analogue system is frequency, temperature, signal chain, component, RFI-environment and pointing angle dependent, and needs to be continuously calibrated on timescales faster than the fastest varying of the set of all these.\\

A first approach to this calibration problem would be the correlation approach. At a minimum, a correlation-based technique must perform an Nx1 correlation. A full correlation of all baselines would allow a over-constrained set of linear equations to be formed, providing a robust solution for gain and phase of each element. In this case, full broadband signal path calibration is possible with any signal, including noise. However, a real-time correlator may require an even larger digital system than the entire beamformer. Thus, correlation approaches may be required to be time multiplexed.\\

Another possible solution is to use the existing beamformer architecture to calibrate off a strong signal. This signal may be externally injected into the analogue system either directly with a noise diode power-combiner scheme or with a known strong signal beamed at the entire array. We chose the latter for our calibration and beam a CW source at the antenna array from the far field.

\subsection{Calibration Procedure}
The calibration procedure for the array for each frequency bin in the band proceeds as follows:

\begin{enumerate}
\item{Fix the correction co-efficient for a reference element}
\item{Sweep the phase of a second element, and record the power output of the beamformer for 2$\pi$ of phase}
\item{Record the coefficient that results in the largest power output of the beamformer, as well as the output power}
\item{Repeat for all elements}
\item{For each element, upload a coefficient $e^{i\phi}$ such as to compensate for the phase error of each element}
\end{enumerate}

\section{Results and Discussion}

\subsection{Anechoic Chamber Testing}
\begin{figure}
\begin{center}
\includegraphics[width=90mm]{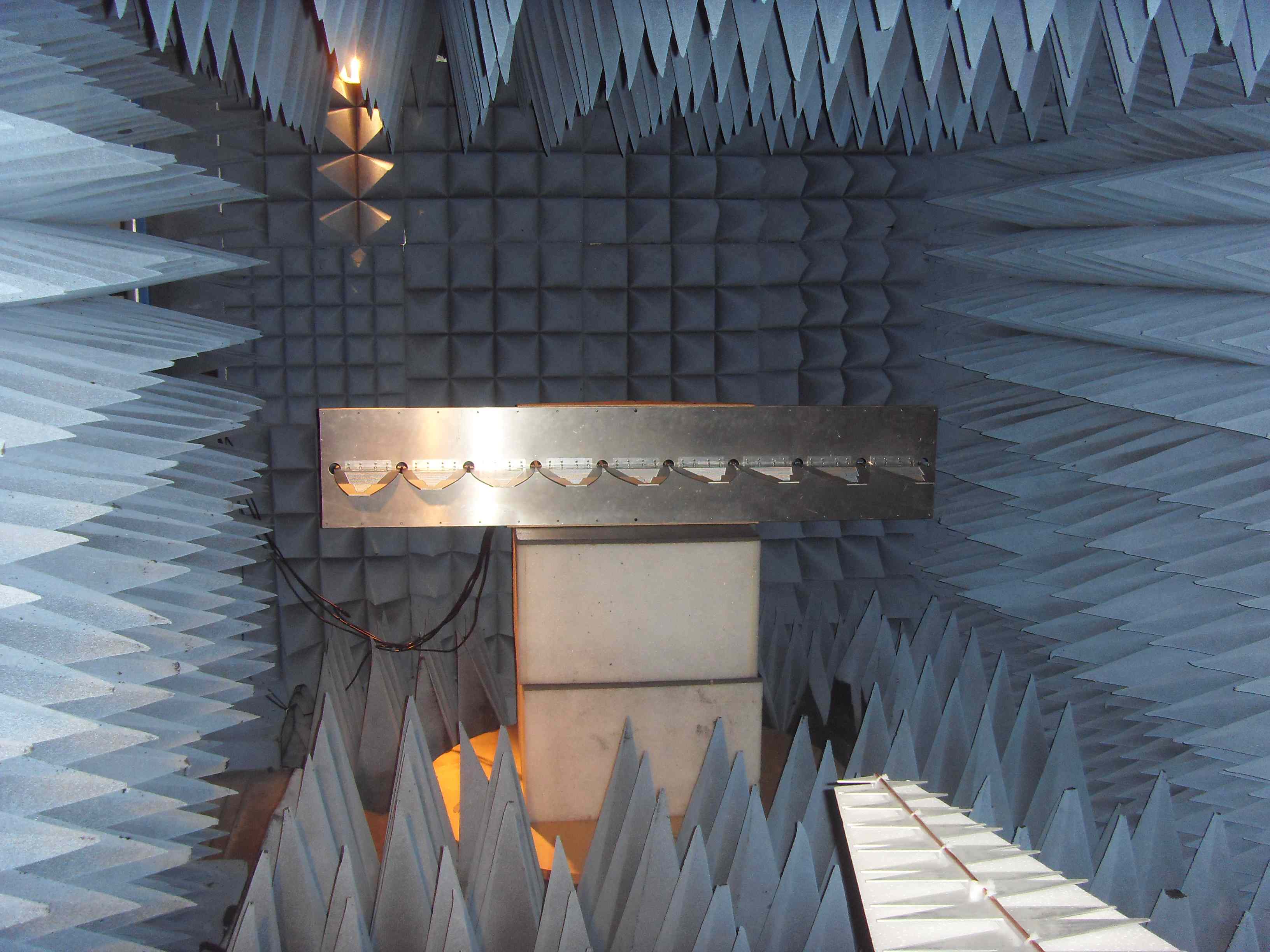}
\caption
	{An image showing the four-element antenna array in the anechoic chamber.} 
\label{AC}
\end{center}
\end{figure}

We have tested the 4-element beamformer using this calibration procedure in a radio-frequency anechoic chamber at the Department of Engineering at Oxford University. Significantly, the beamformer is tested \emph{without} the analogue chain that is required in the eventual radio instrument in the field. Figure \ref{AC} shows the linear, four-element antenna array in the chamber. The transmit antenna is located in the far-field of the receive array so we neglect the transmit aperture illumination function.\\

\begin{figure}
\begin{center}
\includegraphics[width=90mm]{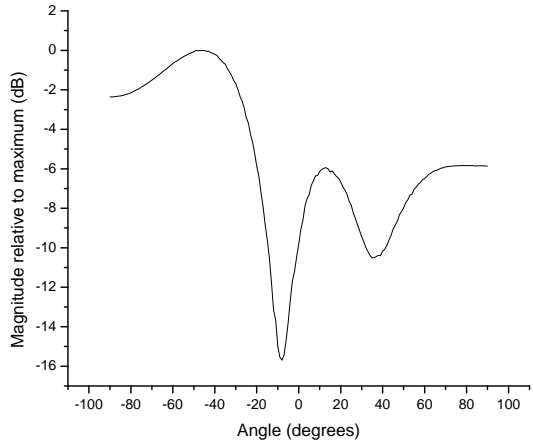}
\caption
	{Beamformer response in the anechoic chamber for a transmitter located at 45 degrees from broadside. The array has been calibrated for the transmitter and a digital beam is swept across the array. The abscissa is the scan angle in degrees from broadside and the ordinate is an arbitrary power scale relative to maximum signal power.} 
\label{45deg digital}
\end{center}
\end{figure}

Figure \ref{45deg digital} shows the beamformer response to a source at 45 degrees from broadside when a digital beam is swept across the aperture.

\subsection{Field Deployment}
\begin{figure}
\begin{center}
\includegraphics[width=90mm]{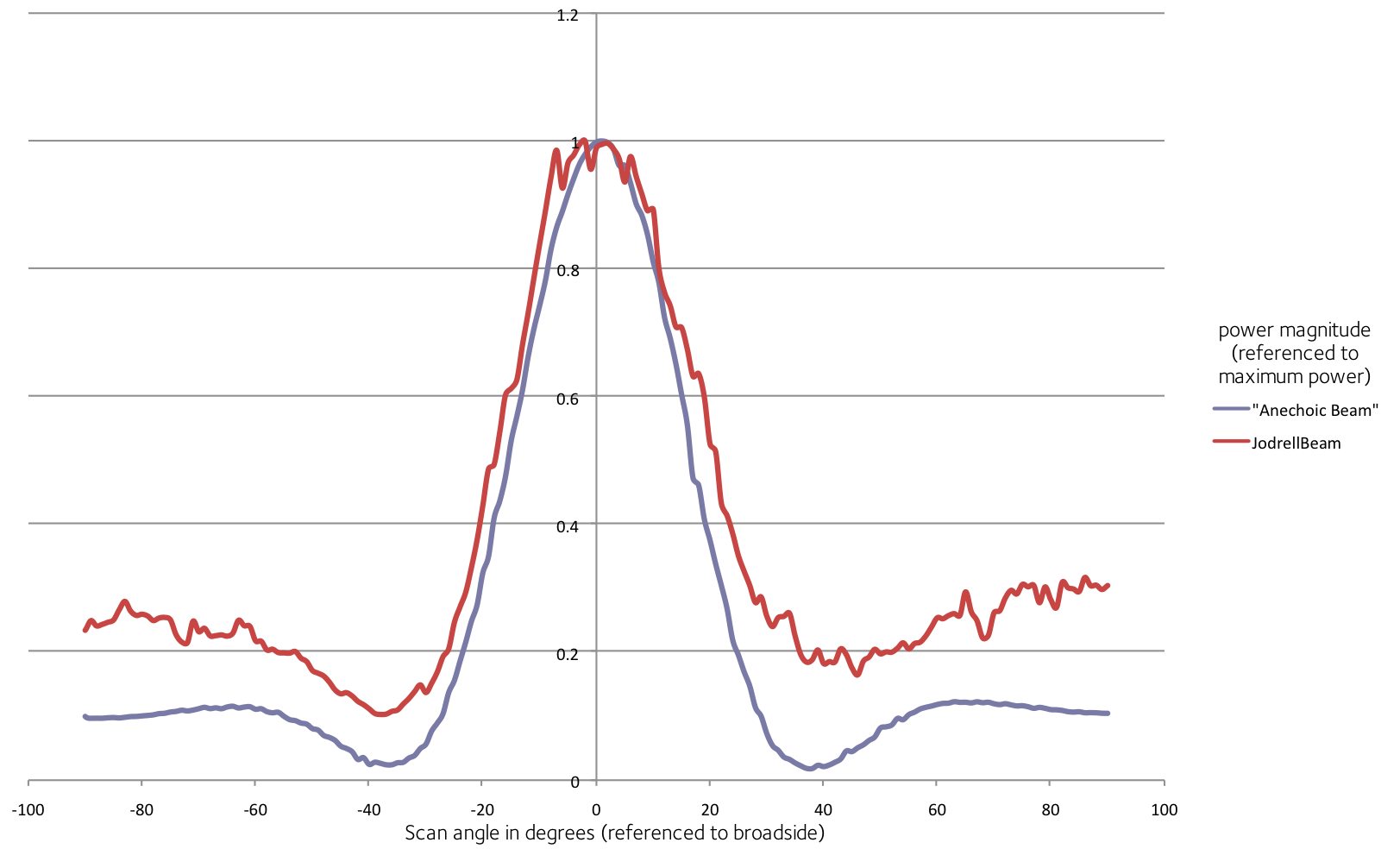}
\caption
	{A plot of the beam pattern of the array at 700MHz in both an ideal RFI environment (anechoic chamber) and in the field. Each of the plots is taken after performing the calibration procedure described earlier. The abscissa is the scan angle in degrees from broadside and the ordinate is an arbitrary power scale relative to maximum signal power at the calibrated 0 degree pointing.} 
\label{ACvsField}
\end{center}
\end{figure}

This calibration process has also been trialled in the field. A transmitter is positioned atop a mast (again, in the far-field of the receive array) and pointed at the array. However, in this case, we included the analogue chain to the antennas before digitisation. The calibration procedure was performed in the same way as before. Figure \ref{ACvsField} shows a 700MHz beam at 0 degrees obtained in the anechoic chamber overlaid with a beam at the same frequency obtained in the field.\\

We expect that the `field' beam will be corrupted by effects like structure scattering\footnote{where structure scattering is taken to mean signal reflection off array structures, ground planes and surrounding vegetation} and the various analogue system effects. Figure \ref{ACvsField} clearly shows that field calibration and measurement results in a beam with significant error compared to the beam obtained in the anechoic chamber.\\

\subsection{Future Work}
The extension to more elements and larger bandwidth is the natural progression for this work. We are currently working on the design of a 16 element, dual polarisation beamformer.\\

The full digital calibration of close-packed, strongly electromagnetically coupled antenna arrays has been shown to be a challenging problem. We intend to develop a more robust calibration scheme in order to achieve the high dynamic range required of digital aperture arrays. Currently, we are investigating full-NxN correlation-based calibration for this task.

\section{Conclusion}
In this paper we propose that digital aperture arrays are a fast, useful and elegant collector technology for GHz radio astronomy, that the development of high-performance signal processing systems is critical to the success of next-generation radio instruments, and have described the design of a 4-element frequency-domain aperture array beamformer.\\





\bibliographystyle{IEEEtran}

%





\label{lastpage}

\end{document}